\documentclass[aps,pre,reprint,superscriptaddress]{revtex4-2}
\usepackage{amsmath,amsfonts,amssymb}
\usepackage{graphicx}
\usepackage{dcolumn}
\usepackage{bm}
\usepackage{hyperref}
\usepackage{xcolor}
\usepackage{comment}

\begin{document}
	    \preprint{APS/123-QED}
	
	\title{Latitudinal dependence of heat transport in  turbulent geostrophic convection}
	
	\author{Veeraraghavan Kannan}
	\affiliation{%
		Max Planck Institute for Solar System Research, Göttingen, 37077, Germany.
	}%
	\author{Varghese Mathai}
	\affiliation{%
	Department of Physics,  University of Massachusetts, Amherst, MA 010003, USA.
    }%
    \affiliation{%
	Department of Mechanical and Industrial Engineering,  University of Massachusetts, Amherst, MA 010003, USA.
	}%
	\author{Xiaojue Zhu}
	\affiliation{%
		Max Planck Institute for Solar System Research, Göttingen, 37077, Germany.
	}%

	\date{\today}
	
	\begin{abstract}
    		\noindent 
Latitudinal variations in turbulent heat flux play a key role in the thermal and magnetic evolution of rapidly rotating planets and stars. Although global spherical-shell simulations have documented such variations, explicit latitude-dependent scaling relations for heat transport have remained elusive. Here we employ the rotating Rayleigh–Bénard convection (RRBC) framework with tilted rotation and gravity axes to model convection at different latitudes $\varphi$ in the geostrophic regime. We derive scaling relations for the latitude dependence of convective length scales $\ell(\varphi)$ and the Nusselt number $Nu(\varphi)$. At high latitudes, the scalings $Nu \sim \sin^{-4/3}\varphi$ (near onset) and $Nu \sim \sin^{-4}\varphi$ (above onset) emerge, while at low latitudes $Nu \sim \cos^{4}\varphi$. These predictions are validated against direct numerical simulations of convection in a spherical shell. The results provide a quantitative framework for regional thermal transport in planetary and stellar interiors and establish a unified interpretation of spherical convection that connects naturally with planar RRBC turbulence.
	\end{abstract}
	
	\maketitle
	\newpage
	
	\section{Introduction}
	
Turbulent convection is fundamental to many geophysical and astrophysical systems, including Earth’s atmosphere, the oceans~\cite{Vallis_Book_2017}, the interiors of stars and planets~\cite{SchumacherS2020, HeimpelGW2016}, and accretion disks surrounding black holes~\cite{AbramowiczLSS1992}. In these systems, large-scale transport is governed by the competing effects of rotation, thermal buoyancy, and latitude~\cite{Chandrasekhar1961}. An accurate description of heat and angular momentum transport in spherical shell convection is critical for understanding the geodynamo, solar cycles, and the thermal evolution of gas giants such as Jupiter. Beyond these, rotating convection is also important in the formulation of climate models, the shaping of global circulation patterns such as Hadley and Polar cells, and the modulation of jet streams and ocean currents~\cite{Houze2004, HeldH1980}. These processes regulate Earth’s radiative balance, precipitation patterns, and climate zones~\cite{ManabeB1985}, underscoring the need for robust models of convection to predict weather and mitigate the impacts of climate change.

\begin{figure}
		\centerline{
			\includegraphics[width=\linewidth]{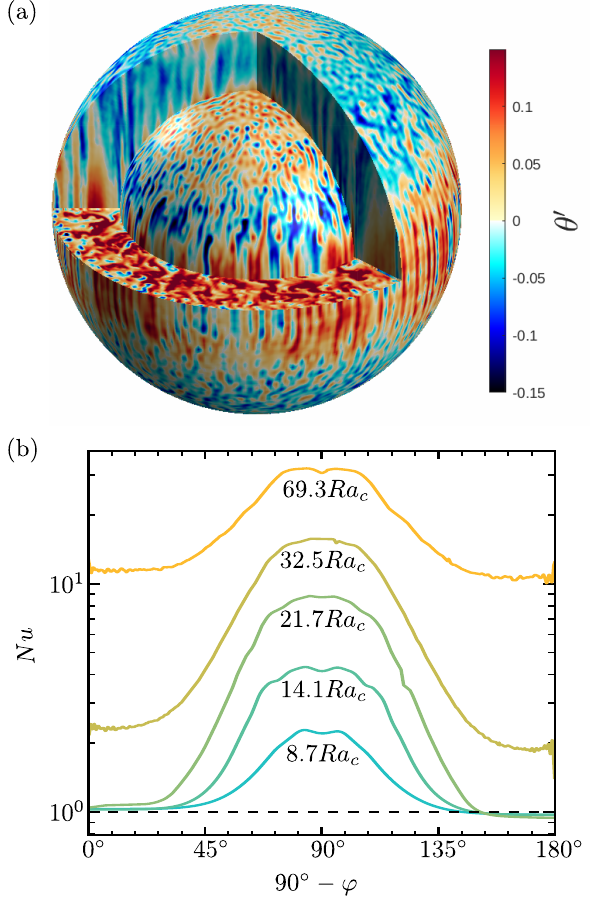}}
		\caption{(a) Contours of the temperature fluctuation $\theta^\prime$ on two meridional cuts, the equatorial section, and two spherical surfaces (corresponding to the inner and outer thermal boundary layers) for $Ek = 10^{-5}$, $Ra = 5 \times 10^7$, $Pr = 1$, radius ratio $\eta = 0.6$, and gravity profile $g(r) \sim (r_o/r)^2$. (b) Time-averaged local Nusselt number $Nu$ at the inner spherical shell boundary as a function of colatitude ($90^\circ - \varphi$) for simulations with $r_i/r_o = 0.6$, $g(r) = (r_o/r)^2$, $Ek = 10^{-6}$, $Pr = 1$, and increasing supercriticalities ($Ra/Ra_c$, where $Ra_c = Ek^{-4/3}$). Data are sourced from \citet{GastineA2023}. }
		\label{fig1}
	\end{figure}

The flows in planetary and stellar interiors are effectively modeled as rapidly rotating, nearly spherical shell geometries. In this regime, Coriolis effects significantly alter turbulent convection, often giving rise to coherent, columnar vortices aligned with the local rotation axis. These structures can produce strongly latitude-dependent flow features~\cite{GastineWA2016, YadavGCDR2016, RaynaudRPGP2018, WangSLV2021}. A sharp contrast is observed between polar convection, dominated by vertically aligned columns, and equatorial convection, shaped by interactions with curved boundaries (see Fig.~\ref{fig1}a). Mid-latitude regions can exhibit distinct regimes, including diffusion-free scaling, in which heat transfer is independent of molecular diffusivity~\cite{WangSLV2021}. As a result, polar, mid-latitude, and equatorial zones exhibit different transport properties~\cite{WangSLV2021, GastineA2023, HartmannSLV2024}. This is illustrated in Fig.~\ref{fig1}b, which shows the variation of the Nusselt number, $Nu \equiv q/q_c$, with colatitude $(90^\circ - \varphi)$, where $q$ is the total vertical heat flux and $q_c$ is the theoretical conductive heat flux. 

These regional variations in flow structures and heat transport evolve as the convective driving increases, transitioning from the \emph{near-onset} to \emph{above-onset} regimes, as depicted in Fig.~\ref{fig1}b. Consequently, the scaling laws governing heat transfer and convective length scales vary significantly with both driving strength and latitude. Global transport models often obscure these localized variations in spherical shell domains~\cite{WangSLV2021, GastineA2023}, and a unifying understanding of the dynamics in latitude-dependent systems is only beginning to emerge~\cite{Novi2019, CurrieBLB2020, TroGJ2024, MiquelEJCK2024}. \citet{GastineA2023} and \citet{WangSLV2021} have recently mapped out a pronounced polar-to-equator contrast in heat flux. These findings establish that latitude is a key control parameter, but most results remain empirical. A predictive theory of how heat transport varies with latitude in spherical shell convection remains elusive.

The rotating Rayleigh--Bénard convection (RRBC) framework serves as a canonical configuration~\cite{EckeS2023, Kunnen2021}, offering valuable insights into processes such as Earth’s magnetic field generation, zonal flows in gas giants, and differential rotation in stars. Localized RRBC models with tilted rotation and gravity axes have been developed to capture the latitude-dependent dynamics observed in spherical shell domains~\cite{Vallis_Book_2017}. The tilted convection approach offers a useful perspective by mimicking dynamics at specific latitudes and has revealed a range of rich phenomena, including large-scale vortices and horizontal jets~\cite{HathawayGT1979, JulienK1998, Novi2019, CurrieBLB2020, WangSLV2021}.

	In this work, we present the first closed-form, latitude-decoupled relations describing the thermal transport in spherical shell convection, spanning from the \emph{near-onset} to the \emph{above-onset} regime. By isolating the dynamics at mid- and low latitudes, we provide insights into the interaction between rotation and buoyancy and their influence on heat transport across latitudes over a broad range of thermal and rotational driving. We compare our predictions with transport properties obtained from direct numerical simulations (DNS) of spherical shell convection and from RRBC configurations with tilted rotation and gravity axes. While, until now, the latitude dependence had been limited to empirical observations, the scalings we derive here successfully explain the polar-to-equatorial contrasts observed in prior work~\cite{GastineA2023, WangSLV2021}. In addition, our work bridges the gap between planar and spherical shell convection, providing a unified modeling perspective on the distinct transport zones in planetary interiors.

\section{Numerical Framework}
We simulate rotating turbulent convection in a Cartesian domain with a rotation vector generally tilted with respect to $z$, where $\zeta$ is the local direction of the rotation vector $\mathbf{\Omega}$ at latitude $\varphi$. Here, varying $\varphi$ from $0^\circ$ to $90^\circ$ corresponds to a change in location from the equator to the pole in spherical geometry. We define $x$ to point eastward, $y$ to point northward, and $z$ to point upward. The geometry of this setup is illustrated in Fig.~\ref{fig2}a. The approach uses a Rayleigh-B\'enard setup where convection is driven by fixed temperature boundary conditions along $z$. The dimensionless governing equations in the rotating Rayleigh-Bénard convection (RRBC) framework with the Boussinesq approximation are:
\begin{equation}
	\partial_t \mathbf{u} + \mathbf{u \cdot} \nabla \mathbf{u} = -\nabla{p} + 
	\sqrt{\frac{Pr}{Ra}} \ \nabla^2{\mathbf{u}} 
	+ T\hat{z} - \frac{1}{Ro}\hat{\zeta} \times \mathbf{u}
	\label{eq1}
\end{equation}
\begin{equation}
	\partial_t \theta + \mathbf{u \cdot} \nabla \theta = \frac{1}{\sqrt{PrRa}} \nabla^2{\theta}
	\label{eq2}
\end{equation}
\begin{equation}
	\nabla \cdot \mathbf{u} = 0
	\label{eq3}
\end{equation}
where $\mathbf{u}$, $p$, and $\theta$ represent the velocity, pressure, and temperature fields, respectively.  In the RRBC setup, the gravity vector $\mathbf{g}$ points vertically downward, while the local background rotation vector $\hat{\zeta}$ is tilted relative to the horizontal $y$-direction, forming an angle $\varphi$. Varying $\varphi$ from $0^\circ$ to $90^\circ$ corresponds to moving from the equator to the pole. Using the height $H$ of the domain as a reference length scale and the buoyancy/free-fall velocity $u_f = \sqrt{\alpha_T \Delta T g H}$ as the reference velocity scale, where $\Delta T$ is the temperature difference between the top and bottom plates and $\alpha_T$ is the thermal expansion coefficient of the fluid, we deduce the non-dimensional control parameters in the RRBC system:
\begin{equation}  \label{eq:parameters}
	\Gamma = \frac{L}{H},\;\; Pr = \frac{\nu}{\kappa},\;\; Ek = \frac{\nu}{2\Omega H^2},
	\;\; Ra = \frac{\alpha_T g \Delta T H^3}{\nu \kappa}
\end{equation}
 Here, $\Gamma$ is the aspect ratio of the system, with $L$ denoting the horizontal length of the domain, $\nu$ is the kinematic viscosity, and $\kappa$ is the thermal diffusivity of the fluid. The Prandtl number $Pr$ measures the ratio of viscous diffusivity $\nu$ to thermal diffusivity $\kappa$ and is a fluid property. The Rayleigh number $Ra$ quantifies the strength of convection relative to diffusion, and the Ekman number $Ek$ measures the strength of viscous diffusion relative to background rotation. We combine these parameters to obtain the convective Rossby number $Ro = \sqrt{Ra/Pr} \, Ek$, which estimates the strength of buoyancy relative to the Coriolis force. Here, $Pr = 1$ was chosen as this aligns with benchmark data that resolve spherical latitude-dependent heat transport \citep{GastineA2023,WangSLV2021}, and is routinely adopted as a first-order proxy for Earth’s liquid outer core \cite{GastineWA2016, YadavGCDR2016, YadavGCWP2016, SchaefferJNF2017, LongMDT2020}.

The simulations were conducted using a second-order, energy-conserving finite-difference code \texttt{AFiD}~\cite{Van2015, Zhu2018}. Various sets of flow parameters were systematically varied by adjusting the latitude $\varphi$ from $10^\circ$ to $90^\circ$ in intervals of $10^\circ$. Time marching was implemented using a third-order Runge-Kutta scheme and the Crank-Nicholson scheme for implicit terms. Boundary conditions included no-slip and constant-temperature conditions at the top and bottom plates, respectively, with periodic boundary conditions applied in the lateral directions. Our simulations differ significantly from previous studies~\cite{HathawayS1983, Novi2019, CurrieBLB2020} due to faster rotation and lower molecular diffusivity compared to the regimes they explored using similar setups. Additionally, we employed a distinct combination of mechanical and thermal boundary conditions in this work. Details of the different cases and the resolution of the numerical grid used in the simulations are listed in Table~\ref{tab01}, along with the time- and volume-averaged response parameters in the RRBC system, namely the Nusselt number, $Nu = \frac{Q}{\kappa \Delta T / H}\equiv 1 + \sqrt{Ra Pr} \langle u_z \theta \rangle_{V,t}$, where $Q$ represents the convective heat flux, $k$ is the thermal conductivity, $u_z$ is the vertical convective velocity, $\theta$ is the temperature deviation, and $\langle \cdot \rangle_{V,t}$ denotes averaging over volume $V$ and time $t$, and characteristic convective length scales from $u_z$ measured along the horizontal $\ell_h$, vertical $\ell_z$, and perpendicular to the local rotational axis $\ell_\perp$.
\begin{table}[!h]
	\caption{Control parameters and numerical results from DNS of tilted RRBC simulations in the present work for $Pr=1$.}~\label{tab01}
	\centering
	\begin{tabular}{cccccc}
		\hline
		$\varphi$ &   $Nu$  & $\ell_h/H$ & $\ell_z/H$ & $\lambda = \ell_h / \ell_z$ & $\ell_\perp/L$ \\ \hline
		\multicolumn{5}{c}{I -- $\Gamma = 4$ \ $Ek = 10^{-5}$ \ $Ra=1\times10^8$ \ $Ra/Ra_c=21.4$} &  \\
		\multicolumn{5}{c}{ $N_x\times N_y\times N_z$ = $512^2\times256$}              &  \\
		10     & $12.67$    &  $1.588$  &  $0.279$  &  $5.686$   &  $0.238$  \\ 
		20     & $11.04$    &  $0.942$  &  $0.275$  &  $3.429$   &  $0.196$  \\ 
		30     &  $8.34$   & $0.576$  &  $0.277$  &  $2.078$   &  $0.162$  \\ 
		40     &  $6.19$   &  $0.422$  &  $0.274$  &  $1.543$   &  $0.126$  \\
		50     &  $4.59$  &  $0.303$  &  $0.306$  &  $0.990$   &  $0.075$  \\ 
		60     &  $3.71$   &  $0.248$  &  $0.314$  &  $0.791$   &  $0.067$  \\ 
		70     &  $3.30$  &  $0.206$  &  $0.402$  &  $0.511$   &  $0.043$  \\ 
		80     &  $3.29$   &  $0.178$  &  $0.628$  &  $0.283$   &  $0.039$  \\ 
		90     &  $3.23$ &  $0.168$  &  $0.879$  &  $0.179$   &  $0.042$  \\[1mm] 
		\multicolumn{6}{c}{II -- $\Gamma = 2$ \ $Ek = 10^{-6}$ \ $Ra=10^9$ \ $Ra/Ra_c=10$}           \\
		\multicolumn{6}{c}{ $N_x\times N_y\times N_z$ = $768^2\times256$}                   \\
		10     & $12.09$  & $0.897$  &  $0.154$  &  $5.810$   &  $0.131$  \\ 
		20     &  $9.47$  &  $0.462$  &  $0.155$  &  $2.966$   &  $0.112$  \\ 
		30     &  $5.25$    &  $0.296$  &  $0.154$  &  $1.921$   &  $0.091$  \\ 
		40     &  $2.89$  &  $0.235$  &  $0.156$  &  $1.509$   &  $0.073$  \\ 
		50     &  $1.94$   &  $0.156$  &  $0.169$  &  $0.919$   &  $0.072$  \\ 
		60     &  $1.50$ &  $0.113$  &  $0.226$  &  $0.499$   &  $0.047$  \\ 
		70     &  $1.23$ &  $0.095$  &  $0.379$  &  $0.249$   &  $0.039$  \\ 
		80     &  $1.14$  &  $0.083$  &  $0.482$  &  $0.172$   &  $0.036$  \\ 
		90     &  $1.11$  &  $0.078$  &  $0.963$  &  $0.082$   &  $0.039$  \\[1mm]
		\multicolumn{6}{c}{III -- $\Gamma = 1$ \ $Ek = 10^{-7}$ \ $Ra=2\times10^{10}$  \ $Ra/Ra_c=9.3$}    \\
		\multicolumn{6}{c}{ $N_x\times N_y\times N_z$ = $768^2\times384$}                   \\
		10     & $12.98$ &  $0.428$  &  $0.086$  &  $4.945$   &  $0.074$  \\
		20     & $11.63$ &  $0.211$  &  $0.087$  &  $2.417$   &  $0.063$  \\ 
		30     &  $5.74$  &  $0.134$  &  $0.086$  &  $1.551$   &  $0.051$  \\ 
		40     &  $3.31$  &  $0.099$  &  $0.086$  &  $1.153$   &  $0.040$  \\
		50     &  $2.17$  &  $0.073$  &  $0.074$  &  $0.986$   &  $0.036$  \\
		60     &  $1.75$  &  $0.054$  &  $0.029$  &  $0.583$   &  $0.036$  \\ 
		70     &  $1.48$   &  $0.043$  &  $0.132$  &  $0.327$   &  $0.035$  \\ 
		80     &  $1.44$  &  $0.038$  &  $0.181$  &  $0.210$   &  $0.033$  \\
		90     &  $1.41$  &  $0.036$  &  $0.971$  &  $0.037$   &  $0.035$  \\[1mm] 
		\hline
	\end{tabular}
\end{table}

\section{Results \& Discussion} 

\begin{figure*}
	\centerline{
		\includegraphics[width=\linewidth]{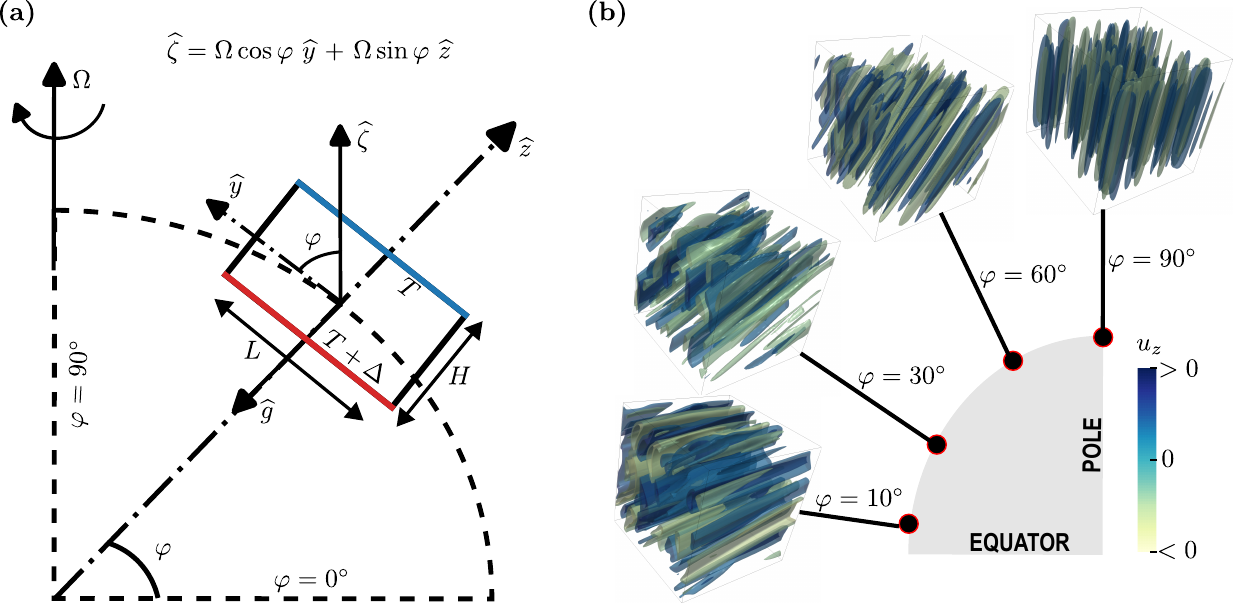}}
	\caption{(a) Schematic illustrating the rotating Rayleigh-B\'{e}nard convection (RRBC) framework, where the rotation vector $\mathbf{\Omega}$ and gravity vector $\mathbf{\hat{g}}$ are misaligned. Here, $\mathbf{\hat{\zeta}}$ represents the local rotation vector for a Cartesian box of height $H$ and length $L$, positioned at latitude $\varphi$. The temperature difference $\Delta$ is defined between the bottom (red) and top (blue) boundaries of the computational domain. (b) Iso-surface plots of vertical velocity $u_z$ in a unit cube at various latitudes, computed for $Ra = 10^9$, $Ek = 10^{-6}$, $\Gamma = 2$, and $Pr = 1$.
	}
	\label{fig2}
\end{figure*}

In the geostrophic regime near-onset where rotation dominates over buoyancy, the flow structures are primarily shaped by rotational effects. Unlike typical convective flows dominated by rolls or plumes, the flow adopts slender convective cells aligned along the rotation axis (see Fig.~\ref{fig2}b). This behavior arises from the Taylor-Proudman balance~\cite{Proudman1916, Taylor1923}, which constrains flow dynamics under rapid rotation. The balance elongates flow features along the rotation axis, rendering the dynamics approximately invariant along this direction. Figure~\ref{fig2}b depicts typical vertical velocity ($u_z$) contours for $Ek = 10^{-6}$, $Ra = 10^9$, $Pr = 1$, and $\Gamma = 2$ during statistically stationary flow at various latitudes. The convective cells consistently align along the local axis of rotation, a trend observed across all simulated latitudes. 

The dominant length scales in the vertical ($\ell_z$) and horizontal ($\ell_h$) directions of the Cartesian domain can be computed from the sinusoidal and Fourier transforms of the vertical velocity $u_z$ along the $z$ and $x$ directions, respectively~\cite{Canuto2007}. Specifically,
\begin{equation*}
	\ell_i/H = \Sigma_{k_i}\left[\hat{u}_z\left(k_i\right)\hat{u}^*_z\left(k_i\right)\right] / \Sigma_{k_i} k_i\left[\hat{u}_z\left(k_i\right)\hat{u}^*_z\left(k_i\right)\right].
\end{equation*}

\begin{figure}[!b]
	\centerline{
		\includegraphics[width=\linewidth]{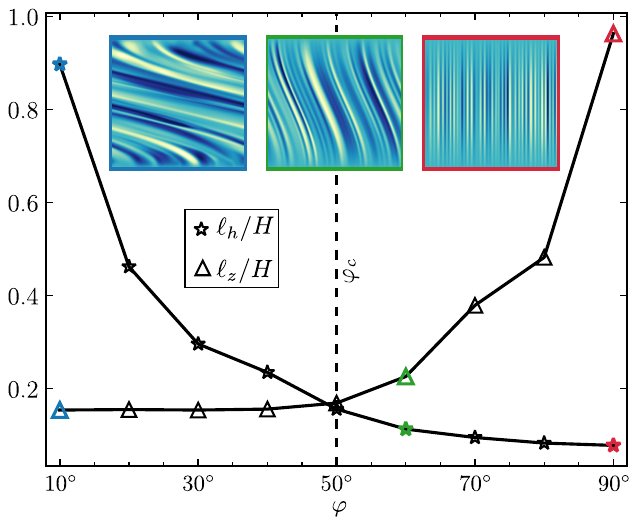}}
	\caption{ Variation of the dominant length scales $\ell_h$ and $\ell_z$, computed in the horizontal and vertical directions, respectively, at different latitudes for $Ra = 10^9$, $Ek = 10^{-6}$, $Pr = 1$, and $\Gamma = 2$. These length scales are obtained from Fourier and sinusoidal transforms of the vertical velocity component $u_z$ along the $x$ and $z$ directions, respectively. A crossover between $\ell_h$ and $\ell_z$ occurs at a critical latitude $\varphi_c \approx 50^\circ$, indicating a transition from vertically to horizontally aligned flow structures.} 
	\label{fig3}
\end{figure}

At the higher latitudes, the convective structure is predominantly oriented along the vertical axis ($\ell_z > \ell_h$), approaching $\ell_z \to H$ as $\varphi \to 90^\circ$, since convective cells tend to align vertically with the rotation axis near the poles, consistent with Taylor–Proudman constraints. Conversely, at lower latitudes, the convective structure is predominantly oriented along the horizontal axis ($\ell_h > \ell_z$), approaching $\ell_h \to L$ as $\varphi \to 0^\circ$. We hypothesize a transitional latitude $\varphi = \varphi_c$, where the convective structure transitions from being vertically dominated to horizontally dominated in scale. This transition results in two distinct flow regimes depending on the ratio $\lambda = \ell_h / \ell_z$: if $\lambda \geq 1$ the flow is horizontally tilted,  and if $\lambda < 1$ the flow is vertically tilted. We use the localized RRBC model to examine these length scales at specific latitudes and quantify the regimes using the ratio of the two length scales. The computed length scales at different latitudes are shown in Fig.\ref{fig3}.  The transition from vertically dominated ($\ell_z > \ell_h$) to horizontally dominated ($\ell_h > \ell_z$) flow structures aligns with our expectation that convective cells tend to align with the local rotation axis, becoming more horizontal at lower latitudes. A similar transition has also been reported in spherical shell simulations\cite{WangSLV2021}.

To understand the latitude scaling of response parameters, it is essential to investigate how the characteristic length scale of the flow structure perpendicular to the local rotation axis, $\ell_\perp$, scales with latitude. 
At high latitudes ($\varphi > \varphi_c$), the flow is rotation-dominant.  In the geostrophic regime near onset, the Viscous-Archimedean-Coriolis (VAC) triple balance is known to hold~\cite{AurnouHJ2020}.  Using a leading-order balance between the vortex stretching by the Coriolis force (perpendicular to the local rotation axis $\zeta$) and vorticity diffusion, we obtain:
\begin{equation*}
	2\Omega \nabla_\zeta u \sim \nu \nabla^2_\perp \mathbf{\omega} \implies 2\Omega \left(\frac{u}{\ell_\zeta}\right) \sim \frac{\nu}{\ell_\perp^2}\left(\frac{u}{\ell_\perp}\right),
\end{equation*}
where $\ell_\zeta$ is the length along the local rotation axis, $\ell_\zeta \approx H/\sin\varphi$. Substituting $\ell_\zeta$ and rearranging the terms, we obtain:
\begin{equation}
	\ell_\perp /H \sim Ek^{1/3}\sin^{-1/3}\varphi. \label{lscale_onset}
\end{equation}

\begin{figure}[!h]
	\centerline{
		\includegraphics[width=\linewidth]{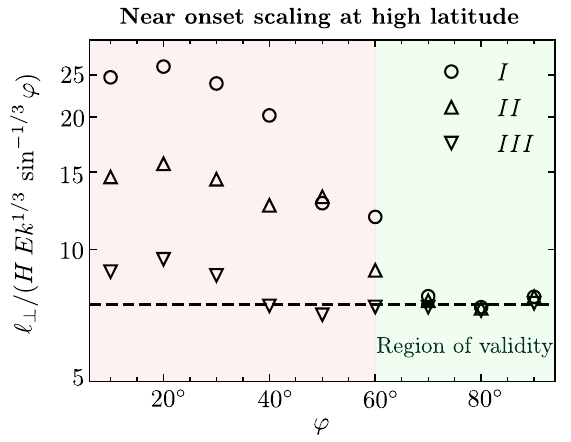}}
	\caption{Verification of the convective length scale $\ell_\perp / H$, as predicted by the scaling laws in Eq.~\eqref{lscale_onset} for near-onset condition. The data are obtained from the present work's DNS simulations of tilted RRBC, presented in Table~\ref{tab01}. Green shaded regions indicate where data fall within the regime of validity of the scaling law, and red shaded regions indicate where data deviate from the scaling law.}
	\label{fig:lscale}
\end{figure}

 The scaling relations for $\ell_\perp$ derived in Eq.~\eqref{lscale_onset} at low latitudes are directly validated using our available Cartesian DNS data (see Fig. ~\ref{fig:lscale}).

Now, the Nusselt number, $Nu$, can be related to $\ell_\perp/H$ in the geostrophic regime (near onset) using the scaling\cite{SongSZ2024}:
$Nu \sim \left(\ell_\perp/H\right)^4 Ra$. This, combined with Eq.\ref{lscale_onset}, yields a latitudinal dependence of $Nu$ for high latitudes (near onset) as,
\begin{equation}
	Nu \sim Ra \ Ek^{4/3}\sin^{-4/3}\varphi \equiv \widetilde{Ra} \sin^{-4/3} \varphi. \label{Nu_onset_high_scale}
\end{equation}

\begin{figure*}
	\centerline{
		\includegraphics[width=\linewidth]{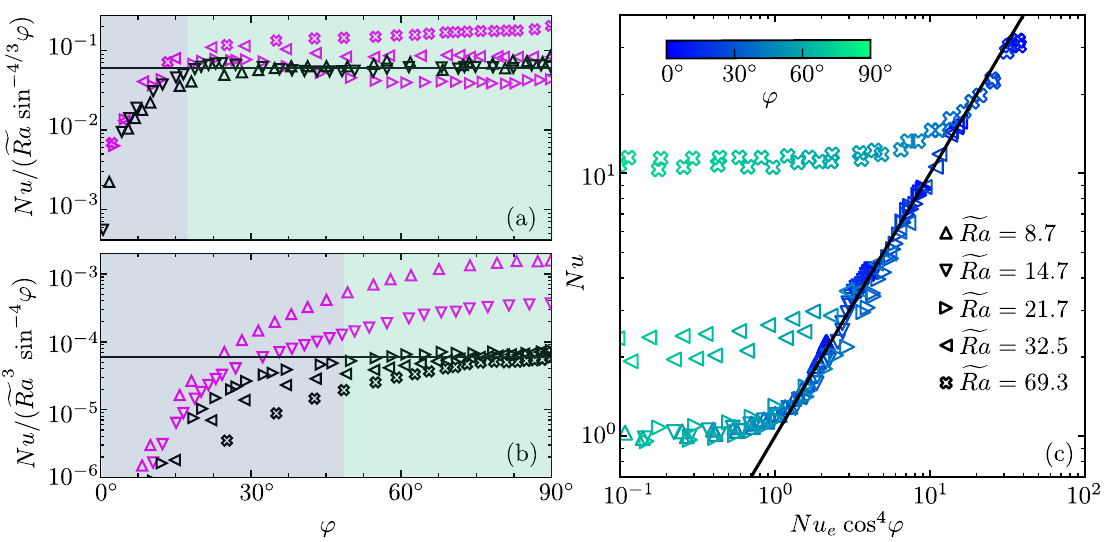}}
	\caption{ Verification of the latitudinal dependence of the Nusselt number $Nu$, as predicted by the scaling laws in Eqs.~\eqref{Nu_onset_high_scale}, \eqref{Nu_turb_high_scale}, and \eqref{Nu_low_scale}, is shown in panels~(a), (b), and (c), respectively.  Panels~(a) and (b) display all five $\widetilde{Ra}$ values, where the black colored datasets correspond to the regime of validity of the proposed scaling laws, i.e. $\widetilde{Ra} <$ 20 for (a) and $\widetilde{Ra} >$ 20 for (b). The purple datasets show a breakdown of the scalings ((a) $\widetilde{Ra} > $ 20   and  (b) $\widetilde{Ra} <$ 20). The data were obtained from direct numerical simulations (DNS) of spherical-shell RRBC, sourced from \citet{GastineA2023}, as presented in Figure~\ref{fig1}.}
	\label{fig4}
\end{figure*}
We note that the geostrophic regime {\it above onset} is not expected to obey the near-onset scaling relation in Eq.~\eqref{Nu_onset_high_scale}. Here, we draw a balance between the Coriolis term along the local rotation axis $\zeta$ and the advection term in the vorticity equation. This balance is justified in the turbulent, rapidly rotating limit of thermal convection~\cite{Ingersoll1982, Aubert2001}, yielding
\begin{equation*}
	u \cdot \nabla_\perp \omega \sim 2\Omega \nabla_\zeta u.
\end{equation*}
Combining a velocity scale $u \sim u_f$, $\ell_\zeta \approx H/\sin\varphi$ and $\omega \sim u/\ell_\perp$, we obtain:
\begin{equation*}
	\frac{\ell_\perp}{H} \sim \frac{u_f}{2\Omega \ell_\perp} \sin^{-1}\varphi.
\end{equation*}
Here, $Ro_\ell = {u_f}/{2\Omega \ell_\perp}$ is the local Rossby number based on the convective length scale. It has been established that $Ro_\ell \sim Ro$ for both slow and rapid rotation limits in RRBC~\cite{AurnouHJ2020}. 
Rearranging terms, we derive:
\begin{equation}
	\ell_\perp / H \sim Ro \sin^{-1}\varphi \sim Ra^{1/2} Ek \sin^{-1}\varphi. \label{lscale_turb}
\end{equation}
Substituting $\ell_\perp /H$ into the relation \cite{SongSZ2024}: $Nu \sim \left(\ell_\perp/H\right)^4 Ra$, we find the scaling for $Nu$ above onset:
\begin{equation}
	Nu \sim Ra^3Pr^{-2}Ek^4\sin^{-4}\varphi \equiv \widetilde{Ra}^3 \sin^{-4} \varphi. \label{Nu_turb_high_scale}
\end{equation}

For the remaining regime of low latitudes and {\it above onset}, $Nu$ can be related to the characteristic convective length scale, i.e., $Nu \sim (\ell_\perp/H)^4 \; Ra$. This relation remains valid regardless of latitude at relatively low $\widetilde{Ra}$, assuming that kinetic energy dissipation follows a diffusion-dependent form, $\epsilon_u \sim \nu u^2 / \ell_\perp^2$~\cite{SongSZ2024}. The horizontal tilting of flow structures at low latitudes leads to the relation $\ell_\perp \approx \ell_z \cos\varphi$, and thus $Nu \sim (\ell_z/H)^4 Ra \cos^4\varphi$. According to the localized model, $\ell_z$ is independent of $\varphi$ when $\lambda \equiv \ell_x / \ell_z \geq 1$ (see Fig.~\ref{fig3} and Table S1). {\color{black} Convection onsets at lower $Ra$ at low latitudes than at high latitudes, as rotation only marginally suppresses the equatorial flow. Data from \citet{GastineA2023} exceeds $Ra_c$ at the poles, reflecting above-onset conditions at low latitudes. Near-onset scaling at low latitudes requires simulations at $Ra \ll Ra_c$ to resolve the behavior of $Nu$. Nevertheless, $Nu \approx Nu_e \cos^4\varphi$ holds above onset, where $Nu_e$ is the equatorial Nusselt number.} The ratio $\ell_z/H$ depends on $Ra$, $Ek$, $Pr$, and $\Gamma$, and sets the proportionality constant. Flow near the equator is especially complex, and $Nu_e$ is known to defy clear power-law scaling~\cite{GastineA2023}.
Nevertheless, since $Nu_e$ is available from the DNS, this allows us to test the latitude dependence of $Nu$ at low latitudes and above onset as: 
\begin{equation}
Nu \approx Nu_e \cos^4\varphi. \label{Nu_low_scale}
\end{equation}

The theoretical predictions for high-latitude scaling {\it near onset}, {\it above onset}, and low-latitude scaling (Eqs. \ref{Nu_onset_high_scale}, \ref{Nu_turb_high_scale}, and \ref{Nu_low_scale}, respectively) are tested by comparing them with data obtained from $Nu$ in spherical shell turbulent convection \cite{GastineA2023} (see also Figure~\ref{fig1}b). {\color{black} The spherical shell DNS data from \citet{GastineA2023} use no-slip velocity boundary conditions at the inner and outer shell boundaries, consistent with our Cartesian DNS setup.}
The {\it near onset} behavior follows Eq.~\eqref{Nu_onset_high_scale} for $\widetilde{Ra} \equiv Ra \, Ek^{4/3} \leq 20$. Beyond this threshold, i.e., for $\widetilde{Ra} \geq 20$, the predictions show excellent agreement with the {\it above onset} scaling of Eq.~\eqref{Nu_turb_high_scale}. The derived scalings exhibit reasonable agreement with data from spherical shell simulations, holding up to $\varphi \geq 25^\circ$ for {\it near onset} scaling and $\varphi \geq 60^\circ$ for {\it above onset} scaling at high latitudes as shown by the green shaded region in Fig.~\ref{fig4}a \& b. At high $\widetilde{Ra}$, especially at mid-latitudes, the dominant balance may shift from Coriolis and inertia to inertia and buoyancy. This could explain the observed deviation from the predicted scaling. Further exploration of this transition will be pursued in future work. Figure~\ref{fig4}c shows $Nu$ normalized by the low-latitude scaling given in Eq.~\eqref{Nu_low_scale}, using spherical shell data obtained from \citet{GastineA2023}. We observe a good agreement at low latitudes for $\varphi \leq 30^\circ$ (blue colored data points). 

\begin{figure}[!h]
	\centerline{
		\includegraphics[width=\linewidth]{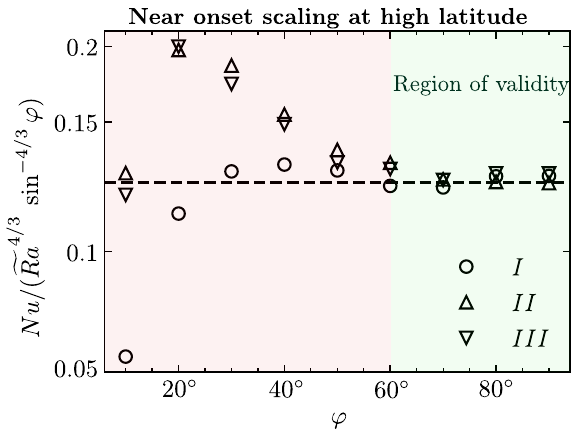}}
	\caption{Verification of the latitudinal dependence of the Nusselt number $Nu$ at high latitudes, as predicted by the scaling law in Eqs.~\eqref{Nu_low_scale} for near-onset condition. The data are obtained from the present work's DNS simulations of tilted RRBC, presented in Table~\ref{tab01}. Green shaded regions indicate where data fall within the regime of validity of the scaling law, and red shaded regions indicate where data deviate from the scaling law.}
	\label{fig:Nuscale}
\end{figure}

To further validate our approach, we compare $Nu$ from available Cartesian DNS  with theoretical scalings (Eqs.~\ref{Nu_onset_high_scale}). There is close agreement at high latitudes near onset (Eq.~\ref{Nu_onset_high_scale}), due to the availability of Cartesian data in this regime, supporting the applicability of the tilted RRBC framework for modeling regional heat transport in spherical geometries (See Fig. ~\ref{fig:Nuscale}). Future work will extend this analysis to regimes well above onset. The condition $\cot \varphi \geq \Gamma$, as noted by \citet{CurrieBLB2020}, ensures that periodic boundary conditions do not constrain low-latitude dynamics, but satisfying it requires larger $\Gamma$ and hence greater computational cost.

\section{Concluding Remarks} 

By establishing an analogy between tilted planar and spherical shell geometries, we developed theoretical predictions for the observed latitude dependence in \textit{spherical shell turbulent convection}---the model geometry commonly used to mimic flows and thermal transport in planetary and stellar interiors. The tilted RRBC framework has enabled us to disentangle the effects of misalignment between rotation and thermal buoyancy. The derived scaling laws---at high latitudes: $Nu \sim \sin^{-4/3}\varphi$ (near onset) and $Nu \sim \sin^{-4}\varphi$ (above onset), and at low latitudes: $Nu \sim \cos^4\varphi$---were validated against global spherical shell turbulence simulations. These findings shed light on latitude-decoupled local heat transport in spherical shell geometries, marking a crucial step toward the development of a comprehensive theory of rotating spherical convection.

Future work will investigate how different mechanical boundary conditions, such as no-slip boundaries (relevant to Earth’s core and icy moons) and stress-free boundaries (applicable to gas giants like Jupiter), influence convective dynamics and heat transport. In particular, no-slip boundaries have been shown to enhance polar heat transport, with important implications for modeling the interiors of icy satellites~\cite{Soderlund2019, HartmannSLV2024}. Understanding how such boundary effects modify flow structures and latitudinal scaling behavior---especially during the transition from geostrophic to non-geostrophic regimes---is essential for connecting idealized models with the diverse conditions found in nature. Another compelling avenue of investigation concerns the influence of the Prandtl number, which differs markedly across planets. A systematic study of these $Pr$-dependent effects lies beyond the scope of the present study and will be undertaken in a separate work.

%

\end{document}